# Comparison of Radon Mass Exhalation Rate Measurements from Building Materials by two Different Methods

Sandra Soares[1, 2,3], Joaquim Kessongo [1, 2,3], Yoenls Bahu[1, 2,3], Luis Peralta[2,4]

[1]Departamento de Física, Faculdade de Ciências da Universidade da Beira Interior, Covilhã, Portugal
[2]Laboratório de Instrumentação e Física Experimental de Partículas (LIP), Lisboa, Portugal
[3]LabExpoRad – UBIMedical, Universidade da Beira Interior, Covilhã, Portugal
[4]Faculdade de Ciências, Universidade de Lisboa, Lisboa, Portugal

**Abstract:** The aim of this study is to measure the radon mass exhalation rate from common granite building materials used in the East and Northeast part of Portugal. Twelve cubic shaped samples were measured. Nine of them without any coating and three coated with different materials (varnish, hydrorepellent and liquid silicone). The radon measurements were performed with two different techniques: one using passive detectors and other using an active detector. For the passive method CR-39 solid state nuclear track detectors were used. The active method used the RAD7 DURRIDGE detector. Radon mass exhalation rates obtained from both methods present relatively low values in the 11 to 45 mBq kg$^{-1}$ h$^{-1}$ range for the analyzed samples. Concerning the coated samples, the measured values are on average four times lower than the ones without coating. Overall, the measured values for both methods present a good agreement.

## 1. Introduction

The human species has always been exposed to ionizing radiation of natural origin, which can be found in soils, some rocks and water. From the late nineteenth century onwards, so-called artificial radiation was added to this background radiation. About 80% of the background radiation arise from natural sources, in which we can include the naturally occurring radioactive materials (NORM) present in the Earth's crust[1,2], in different types of food, beverages and in some building materials. This radiation is largely due to primordial radionuclides of the natural radioactive series of $^{232}$Th and uranium isotopes $^{238}$U; $^{235}$U, and their progeny. A sizable contribution is also given by the radioisotope $^{40}$K. In terms of human health and environment effects the radionuclide with major radiologic interest is radon ($^{222}$Rn)[3]. This radioactive isotope results from the disintegration of $^{226}$Ra, a decay product of the $^{238}$U series and responsible for the largest source of natural radiation to which population is subject, contributing with approximately 50% for the total dose of radiation[4]. Radon concentration can reach high levels in dwellings depending not only on exhalation from soil but also on the building material used like concrete, bricks, phosphogypsum or granite[5]. In general, either to improve waterproofness or indoor performance namely comfort and desegregation of the natural stone limiting the particle loss, some coatings are normally used over natural stone. These coatings can also be used as a way of impermeabilization to radon exhalation. There are several parameters to assess, in a satisfactory way, the levels of airborne radon resulting from soil beneath the house or from building materials. Radon mass exhalation rate is one of these parameters commonly used to express indoor radon released from building and ornamental materials. This quantity can be obtained from the radon concentration in air. The measurement techniques are thus derived from the ones used for radon concentration measurement. Passive detectors or active detector can be employed for this task. The focus of this work is thus to compare the radon exhalation rate from granite stone used as building using two different measurement methods.

## 2. Methods

Inside dwellings the concentrations of radon can usually vary with temperature, humidity, ventilation, building materials and type of house. The radon concentration in building materials can be measured using either passive or active detectors. In this work radon mass exhalation rates for 12 granite samples (figure 1), with and without coating were determined using a passive measuring technique or a fast-electronic measuring technique. The employed coatings are normally used over natural stone and were varnish, hydrorepellent and liquid silicone.

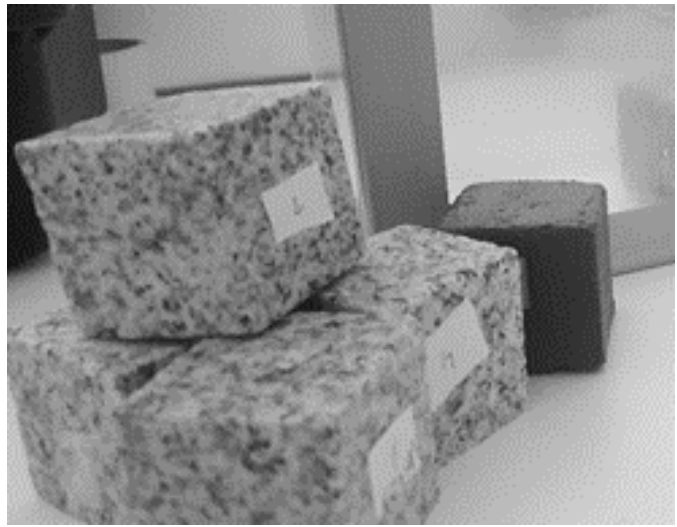

Figure 1. Example of the granite samples used in the study.

### 2.1 Integration detector

In this work the measuring techniques use the Closed-Can method[6,7] This method consists in placing the sample to be measured inside a sealed chamber. All radon exhaled from the sample stays inside the chamber (assuming no-leakage). If N(t) is the number of radon atoms inside the box of volume V at a certain time t, the radon concentration c(t) can be defined as[8,9]

$$c(t) = \frac{\lambda N(t)}{V} \text{ (Bq m}^{-3}\text{)} \qquad (1)$$

where λ is the radon decay constant.

Considering a sample of mass M exhaling radon at a rate that its contribution to the amount of radon concentration in the chamber is $E_R$ (Bq kg$^{-1}$ s$^{-1}$)×M(kg) the time evolution of the concentration in the chamber is given by[3,4]

$$\frac{dc(t)}{dt} = \frac{E_R M}{V} - \lambda c(t) \qquad (2)$$

Assuming no initial activity in the chamber c(0)=0 the solution of the equation is

$$c(t) = \frac{E_R M}{V \lambda}(1 - e^{-\lambda t}). \qquad (3)$$

If the radon gas is uniformly distributed inside the chamber, the number of hits per unit time on the surface of a detector is proportional to the radon concentration

$$\frac{dh(t)}{dt} = \varepsilon c(t) \qquad (4)$$

where h(t) is the recorded number of hits at time t and ε is the total detection efficiency. Integrating the equation over an exposure time T, one obtains the total number of hits on the detector as

$$h(T) = \frac{\varepsilon E_R M}{\lambda V}\left[T - \frac{1}{\lambda}(1 - e^{-\lambda T})\right] \qquad (5)$$

where the quantity in square brackets is often called effective exposure time

$$T_{eff} = \left[T - \frac{1}{\lambda}(1 - e^{-\lambda T})\right]. \qquad (6)$$

Equation 5 can be solved for $E_R$ and written as

$$E_R = \frac{\lambda V h(T)}{\varepsilon M T_{eff}}. \qquad (7)$$

On the other hand, h(T)/ε is the integrated radon concentration over the exposure time T. The average radon concentration $c_R$= h(T)/εT (Bq m$^{-3}$) is the quantity experimentally obtained so that the mass exhalation rate $E_R$ is computed as

$$E_R = \frac{\lambda V c_R T}{M T_{eff}} \quad \text{(Bq kg}^{-1}\text{ h}^{-1}\text{)} \qquad (8)$$

where λ is the $^{222}$Rn decay constant, $\lambda_{Rn222}$=7.554×10$^{-3}$ h$^{-1}$ and the variables T and $T_{eff}$ are given in hour.

If the radon concentration is not zero when the chamber is closed then an additional term must be added to equation 3, describing the decay of the environmental radon pre-existing in the chamber

$$c(t) = \frac{E_R M}{V\lambda}\left(1 - e^{-\lambda t}\right) + c_0 e^{-\lambda t} \qquad (9)$$

where $c_0$ is the existing radon concentration at t=0. The total number of hits is then given by

$$h(T) = \frac{\varepsilon E_R M T_{eff}}{\lambda V} + \frac{\varepsilon c_0 (1 - e^{-\lambda t})}{\lambda}. \qquad (10)$$

As before we relate the total number of hits on the detector with the average radon concentration $h(T) = \varepsilon C_R T$ obtaining for the mass exhalation rate

$$E_R = \frac{\lambda V [c_R T - c_0/\lambda (1 - e^{-\lambda T})]}{M T_{eff}}. \qquad (11)$$

As pointed out by several authors[10,11] the chamber might leak some radon out. In this case the effect is a modification of the decay constant $\lambda$[11] incorporating a leaking term $\lambda_{leak}$ in such a way that $\lambda = \lambda_{Rn222} + \lambda_{leak}$.

## 2.2 Chamber leakage assessment

In order to assess the radon leakage a chamber was prepared to measure the radon concentration in real time. Two essays were performed. One essay using an uranium ore rock and a second essay placing three of the analysed samples in the chamber. In each essay the RadonEye[12] active detector radon source was placed inside the chamber along with the radon sources. Special care was taken to prevent radon from leaking out through the connector cord hole, by sealing it with UHU Patafix® adhesive. Radon concentration in the chamber was allowed to build-up for several days. The measured curves (figure 2) were fitted with the function $c(t) = c_{max}(1 - e^{-\lambda t}) + c_0$. Values of λ=0.0161±0.0002 h⁻¹ (χ²/ν=0.72) and λ=0.0120±0.0003 h⁻¹ (χ²/ν=0.58) for the total decay constant were obtained, and thus values of λ_leak=0.0085±0.0002 h⁻¹ and λ_leak=0.0044±0.0003 h⁻¹ were respectively obtained for the leakage constant. From these results we conclude that leakage has a dependence on the radon concentration gradient between the chamber and the outside. For the lower radon concentration obtained with each granite sample, lower leakage is to be expected. Since the second essay was performed using three samples inside the chamber the value of λ_leak=0.0044±0.0003 h⁻¹ was used in the following analysis as an upper limit, closer to the actual value in each sample case. The existence of leakage has other side effect. If we consider an empty leaky chamber its radon concentration time evolution will not decrease exponentially due to radon exchange with the exterior. From the experimental point of view is then easier to obtain the integrated background concentration $c_{BGD}$ and subtract it from the value obtain for a given sample. The mass the mass exhalation rate equation 11 becomes

$$E_R = \frac{\lambda V (c_R - c_{BGD}) T}{M T_{eff}}. \qquad (12)$$

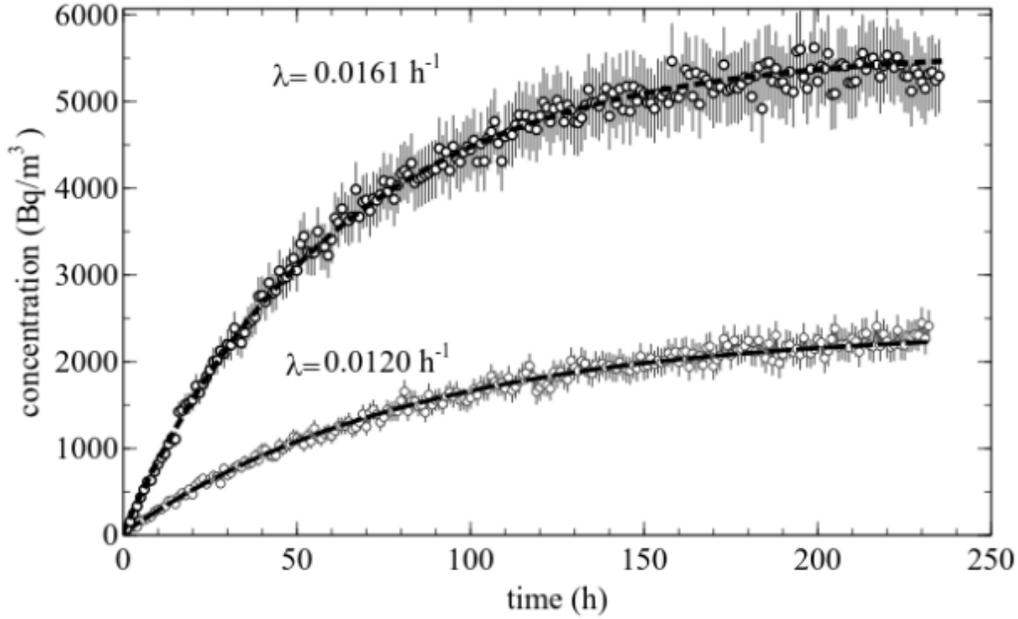

Figure 2: Radon build-up inside the chamber for several days for two different radon sources. Higher leakage constant is obtained for the higher exhalation rate source. The dashed lines are fits to the data. The decay constant obtained by the fit is given.

## 2.3 Differential detectors

Active detectors are capable to give real time radon concentration values (integrated in time intervals much smaller than radon half-life). The measured quantity is thus the radon concentration at a time t given by equation 9, where $\lambda$ is the total decay constant including leaking. Solving the equation for $E_R$ and assuming a background concentration $c_{BGD}$ for the leaky chamber, one obtains for an acquisition time t=T

$$E_R = \frac{\lambda V(c_R(T)-c_{BGD})}{M(1-e^{-\lambda T})} \quad . \quad (13)$$

## 2.4 Passive detector method

This technique consists in placing the granite sample inside an acrylic chamber along with the passive detector. The chamber is then sealed with silicone sealant. In this work CR-39 integration detectors were used. Each CR-39 detector exposure chamber was fixed onto the top center of the can by means of adhesive tape, as shown in figure 3.

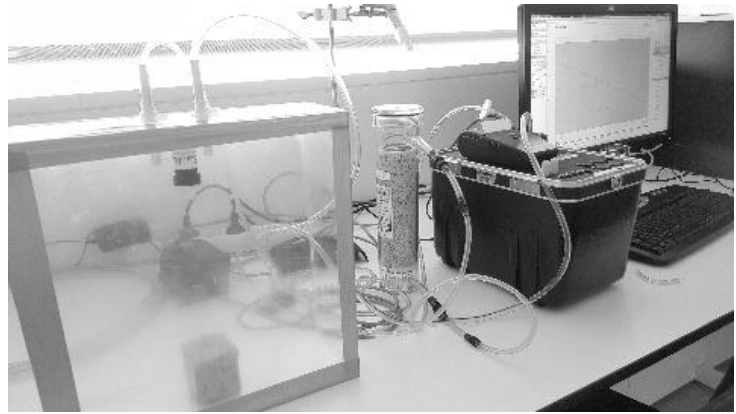

Figure 3. The sealed-can technique for measuring radon exhaled from granite samples with a CR-39 alpha-particle sensitive track detector.

After an exposure time of one week the CR-39 detectors were removed from the chamber and etched chemically in a 6.25 M NaOH solution to display and enlarge the latent alpha tracks due to radon decay. The number of etched alpha tracks were automatically counted using a Radosys NanoReader Track Counting Microscope.

After the determination of the radon concentration inside the chamber, the mass exhalation rate $E_R$ was calculated using equation 12. The volume V is the difference between the volume of the chamber and the volume of the sample. It is assumed radon also fills the CR-39 exposure chamber. For background assessment, radon concentration was measured for an empty chamber following the same procedure as for the stone samples.

## 2.5 Active detector method

The radon concentration was measured with a RAD7 detector[13]. This device has an alpha detector that measures the signals from the radon progeny, namely the $^{218}$Po and $^{214}$Po isotopes. Mass exhalation rates from investigated samples, were estimated by placing the samples in a chamber with a closed loop arrangement: the chamber was connected through two vents to the inlet and outlet of the RAD7 device (figure 4).

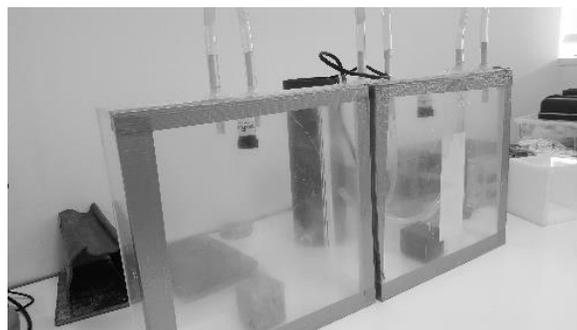

Figure 4. The RAD7 device for measuring radon concentration of granite samples.

The chamber volume and expose time T were identical to the passive method. The radon mass exhalation rate was then obtained with equation 13. The background concentration was obtained measuring the radon concentration in an empty chamber after being closed for the same time T as in the sample measurement.

## 3. Results and discussion

The characteristics of the 12 measured samples are summarized in table 1. The samples were cut in cubes of side 5.0 cm and average mass of 0.330 kg. The values of radon concentration inside the chamber after one week of exposure obtained with the passive and active methods are also presented.

Table 1. Radon concentration measured for the analyzed samples.

| Sample | Coating type | Passive method Radon concentration (Bq/m$^3$) | Active method Radon concentration (Bq/m$^3$) |
|---|---|---|---|
| 1 | No coating | 85±13 | 76±11 |
| 2 | No coating | 33±5 | 29±4 |
| 3 | No coating | 25±4 | 20±3 |
| 4 | No coating | 30±5 | 24±4 |
| 5 | No coating | 32±5 | 28.±4 |
| 6 | No coating | 61±9 | 59±9 |
| 7 | No coating | 48±7 | 44±7 |
| 8 | No coating | 40±6 | 34±5 |
| 9 | No coating | 50±8 | 48±7 |
| 10 | Varnish | 25±4 | 21±3 |
| 11 | Liquid Silicone | 22±3 | 21±3 |
| 12 | Hydrorepellent | 23±4 | 22±3 |

Having determined the radon concentration, the radon exhalation rate was estimated accordingly for both methods. The values are given in figure 5. Preliminary results show that radon mass exhalation rates from the analysed granitic samples have in general relatively low values.

For the passive method, the higher values found was 44.5 mBq kg$^{-1}$ h$^{-1}$, from sample S1. For the coated samples values of 13.1, 11.1 and 11.5 mBq kg$^{-1}$ h$^{-1}$ were obtained respectively for varnish, liquid silicone and hydrorepellent. A factor of 4 is observed between the higher and lower exhalation rates, for the tested granites.

For the active method, the higher value found was 26.2 mBq kg$^{-1}$ h$^{-1}$, from sample S1. This type of stone was chosen to compare the mass exhalation rate of the non-coated and coated samples. For the coated samples values of 7.3, 7.1 and 7.3 mBq kg$^{-1}$ h$^{-1}$ were obtained respectively for varnish, liquid silicone and hydrorepellent.

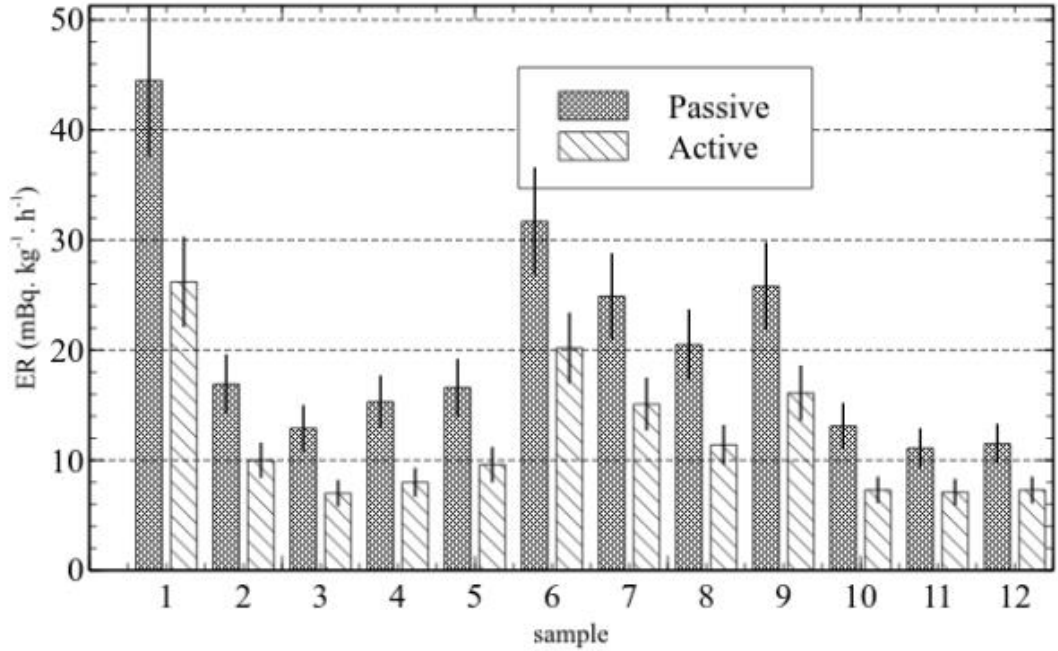

Figure 5. Comparison between the radon mass exhalation rate determined by the passive and active methods for the 12 studied granite samples

## 4. Conclusion

The obtained results show systematically higher values for the passive method. Other authors[10] have reported similar effects. A possible cause is the presence of thoron gas ($^{220}$Rn) which is not separate from $^{222}$Rn by passive detector, unlike active detectors relying the concentration measurement on discriminated alpha peaks intensities. It is clear that measurement conditions are of paramount importance and must be improved, namely the type of chamber used, its sealant and overall size. In this experiment acquisition time was for weak. For passive detectors this is a short time, especially due to their low detection efficiency. Longer acquisition time is advisable to increase the statistical significance of the result. Other important aspect emerging in our work is the importance of radon leakage. Although a considerable amount of care was taken to prevent leakage, still a sizeable fraction of radon escapes the container, leading to an important leakage constant. Our assessment is that the container design should be careful reviewed. The used materials more thoroughly tested, in particular any material used in the junctions. The measurement of radon concentration build-up inside the container, using a real-time acquisition, active detector, is of paramount importance to assess the radon leakage from the container. When making the measurement with passive detectors it is also important to use at least two detectors: one at the bottom and one at the top of the container. In this way any eventual radon concentration gradient can be assessed and, an average value obtained. Finally, the sample should not be in contact with any of the container's surface, but rather suspended. In this way exhalation is equivalent from any of the sample faces. More samples from different materials will be analyzed in the future and the comparison between the two methods made for other materials.